\documentclass[twocolumn]{aastex7}

\newcommand{\kms}          {\mbox{${\rm km~s^{-1}}$}}

\newcommand{\e}            {\mbox{$^{-1}$}}

\def\cm2{\mbox{${\rm cm^{-2}}$}}
\def\h2{\mbox{${\rm H}_2$}}
\def\nh2{\mbox{$n_{\rm H_2}$}}
\def\Nh2{\mbox{$N_{{\rm H}_2}$}}
\def\Mh2{\mbox{$M_{{\rm H}_2}$}}

\def\farcs{\hbox{$.\!\!^{''}$}}

\def\simgt{\lower.5ex\hbox{$\; \buildrel > \over \sim \;$}}
\def\simlt{\lower.5ex\hbox{$\; \buildrel < \over \sim \;$}}

\def\13CO{$^{13}$CO}
\def\C18O{C$^{18}$O}
\def\H2{H$_2$}

\def\Rtilt{\mbox{$R_{\rm tilt}$}}
\def\itilt{\mbox{$i_{\rm tilt}$}}
\def\PAtilt{\mbox{${\rm PA}_{\rm tilt}$}}
\usepackage{amsmath}
\received{Summer 2025}
\accepted{\today}
\submitjournal{ApJ}

\shorttitle{Modeling of a Shadowed Disk}
\shortauthors{Williams et al.}
\begin{document}

\title{Radiative Transfer Modeling of a Shadowed Protoplanetary Disk assisted by a Neural Network}

\correspondingauthor{Jonathan Williams}
\email{jw@hawaii.edu}

\author[orcid=0000-0001-5058-695X]{Jonathan P. Williams}
\affiliation{Institute for Astronomy, University of Hawai'i at Mānoa, 2680 Woodlawn Drive, Honolulu, HI 96822, USA}
\email{jw@hawaii.edu}

\author[orcid=0000-0002-7695-7605]{Myriam Benisty}
\affiliation{Max-Planck Institute for Astronomy (MPIA), Königstuhl 17, 69117 Heidelberg, Germany}
\email{benisty@mpia.de}

\author[orcid=0000-0002-4438-1971]{Christian Ginski}
\affiliation{School of Natural Sciences, Center for Astronomy, University of
Galway, Galway H91 CF50, Ireland}
\email{christian.ginski@universityofgalway.ie}

\author[orcid=0000-0002-2357-7692]{Giuseppe Lodato}
\affiliation{Dipartimento di Fisica, Università degli Studi di Milano, Via Celoria 16, I-20133 Milano, Italy}
\email{giuseppe.lodato@unimi.it}

\author[orcid=0000-0001-5763-378X]{Maria Vincent}
\affiliation{Institute for Astronomy, University of Hawai'i, 640 N. Aohoku Pl., Hilo, HI 96720, USA}
\email{mariavin@hawaii.edu}

\begin{abstract}
We present observations and detailed modeling of a protoplanetary disk around the T Tauri star, V1098 Sco. Millimeter wavelength data from the Atacama Large Millimeter Array (ALMA) show a ring of large dust grains with a central cavity that is filled with molecular gas. Near-infrared data with the Very Large Telescope (VLT) detect the scattered starlight from the disk surface and reveal a large shadow that extends over it's entire southern half. We model the ALMA continuum and line data to determine the outer disk geometry and the central stellar mass. Using radiative transfer models, we demonstrate that a misaligned inner disk, tilted in both inclination and position angle with respect to the outer disk, can reproduce the salient scattered light features seen with the VLT. Applying an image threshold algorithm to compare disk morphologies and training a neural network on a set of high signal-to-noise models, we forward model the data and determine the inner disk geometry. We find that the rotation axes of the inner and outer disks are misaligned by $38^\circ$ and constrain the mass and location of a perturbing planetary or substellar companion. The technique of simulation based inference that is illustrated here is broadly applicable for radiative transfer modeling of other objects.
\end{abstract}


\section{Introduction}
\label{sec:introduction}
The varied orbital architectures of the multitude of planetary systems discovered to date is due to a combination of their formation properties and subsequent interaction with the surrounding disk, together with longer term dynamical effects with the host star and planetary siblings.
Disk properties determine the location of the initial seeds which then symbiotically evolve through gravitational torques that drive gas accretion, gap clearing, and planetary migration.

Planetary embryos form in the dense, flat disk midplane but fully-formed planets can, in some cases, orbit at high inclinations relative to the stellar rotation axis \citep{2018haex.bookE...2T}. This may be caused after the disk has dispersed by planet-planet scattering \citep{2008ApJ...686..621F} or interactions with a companion star via the Kozai mechanism \citep{2003ApJ...589..605W}.
However, orbital obliquity might also be imprinted during formation \citep{2010MNRAS.401.1505B} as some protoplanetary disks (defined as disks around optically visible pre-main sequence stars) also show indications for early evolution of orbital inclination through asymmetric shadows cast by a misaligned inner disk on scattered light from the outer disk \citep{2015ApJ...798L..44M}, from direct imaging of edge-on disks \citep{2024ApJ...961...95V},
and as dips in the starlight from dust along the line of sight toward systems that appear more face-on \citep{2020MNRAS.492..572A}.

For over a decade now, we have had the tools to image protoplanetary disk structure at the scales of planet-disk interactions. The Atacama Large Millimeter Array (ALMA) and the SPHERE instrument on the Very Large Telescope (VLT) have been at the forefront of such work \citep{2023ASPC..534..605B, 2023ASPC..534..423B}.
Both resolve disk structures at scales of a few au in nearby disks and complement each other as the long wavelengths of ALMA reveal the cold molecular gas and relatively large (approximately millimeter-sized) dust grains in the disk midplane whereas SPHERE detects the near-infrared starlight scattered by much smaller (approximately micron-sized) grains in the upper disk atmosphere \citep{2020ARA&A..58..483A}.

In this paper, we present ALMA and SPHERE images of the protoplanetary disk around the T Tauri V1098\,Sco (2MASS J16140792-1938292). V1098\,Sco is of spectral type K1 and lies at a distance of 160\,pc \citep{2020yCat.1350....0G} in the Upper Sco region of the Sco-Cen OB association where the median age is estimated to be $5\pm 2$\,Myr \citep{2016MNRAS.461..794P}.
It was first recognized as a T Tauri star by \citet{1992AJ....103..549G} who carried out a spectroscopic survey of color selected IRAS sources which showed it to have weak and variable H$\alpha$ indicative of low accretion and moderate Lithium indicative of youth.
More recently, it was classified as a quasi-periodic dipper from the analysis of K2 light curves by \citet{2018AJ....156...71C}.
However, this otherwise unremarkable star has a quite striking appearance in scattered light as one half of the disk appears to be missing due to a large shadow and millimeter wavelength imaging reveals a large central cavity.

We present the observations in \S\ref{sec:observations}, fit the ALMA data in \S\ref{sec:outer_disk}, and model the SPHERE scattered light image in \S\ref{sec:radmc}. We carry out a large grid of radiative transfer models and employ a novel method using a neural network to interpolate model images for parameter values between grid points which allows us to efficiently fit the data using an MCMC approach. We find that the outer disk shadow is caused by a large, misaligned inner disk and discuss mechanisms for this particular geometry in \S\ref{sec:discussion}. We then summarize our results in \S\ref{sec:summary}.

\section{Observations}
\label{sec:observations}
\subsection{SPHERE near-infrared observations}

V\,1098\,Sco was observed as part of the ESO DESTINYS (Disk Evolution Study Through Imaging of Nearby Young Stars, \citealt{2020A&A...642A.119G}; project id 1104.C-0415(D)) large program on 30-07-2022 using the IRDIS sub-system of SPHERE. Observations were obtained in DPI mode (dual polarization imagings; \citealt{2020A&A...633A..64V}) in the broad band H filter with an effective wavelength of 1.6$\mu m$. The central star was obscured by a coronagraphic mask with an effective inner working angle (defined as 50\,\% drop in throughput) of 92.5\,milli-arcseconds. The weather conditions during the observations were excellent with optical seeing between $0\farcs4-0\farcs8$ and and a coherence time of the atmosphere between 5\,ms and 11\,ms. We recorded a total of 26 full polarimetric cycles, each comprised of 4 images with different half wave plate orientations as described in \cite{2020A&A...642A.119G}. Each image had an individual exposure time of 32\,s, giving us a total integration time of 55.5\,min. In addition to the science data we recorded dedicated flux, center and sky calibration frames at the beginning and end of the science sequence. \\
The data was reduced using the public IRDAP (IRDIS Data reduction for Accurate Polarimetry) pipeline\footnote{https://irdap.readthedocs.io/en/latest/}. To briefly summarize IRDAP performed first general image processing tasks such as, sky subtraction, flat fielding, and bad pixel correction before each frame was centered using the stellar coordinates measured in the center calibration frames. Using center calibration observations at the beginning and end of the science sequence we find that the determined stellar center position deviates by less than 0.05\,pixel between these two, indicating excellent stability of the stellar position behind the coronagraphic mask. After the initial data processing IRDAP then performed polarization differential imaging which cancels the unpolarized stellar light and retains the partially linearly polarized scattered light from the circumstellar disk. IRDAP accounts for instrumental polarization by means of a detailed instrument model described in \cite{2020A&A...633A..64V}.


\subsection{ALMA mm observations}
We downloaded ALMA observations of V1098\,Sco from the archive \citep[project 2018.1.00564.S, recently published in ][]{2025ApJ...978..117C} and reduced it using the same procedures\footnote{https://github.com/jjtobin/edisk} as for the eDisk project \citep{2023ApJ...951....8O}.
The final data products are Band 7 (340\,GHz) continuum and CO 3--2 spectral line.
The beam and rms noise levels are $0\farcs41 \times 0\farcs32$ at a position angle of $-76^\circ$ and 0.13\,mJy\,beam$^{-1}$ for the continuum and $0\farcs38 \times 0\farcs31$ at $-75^\circ$ and  5\,mJy\,beam$^{-1}$ per 0.5\,\kms\ channel for the line.

\begin{figure*}[ht!]
\begin{center}
\includegraphics[width=\textwidth]{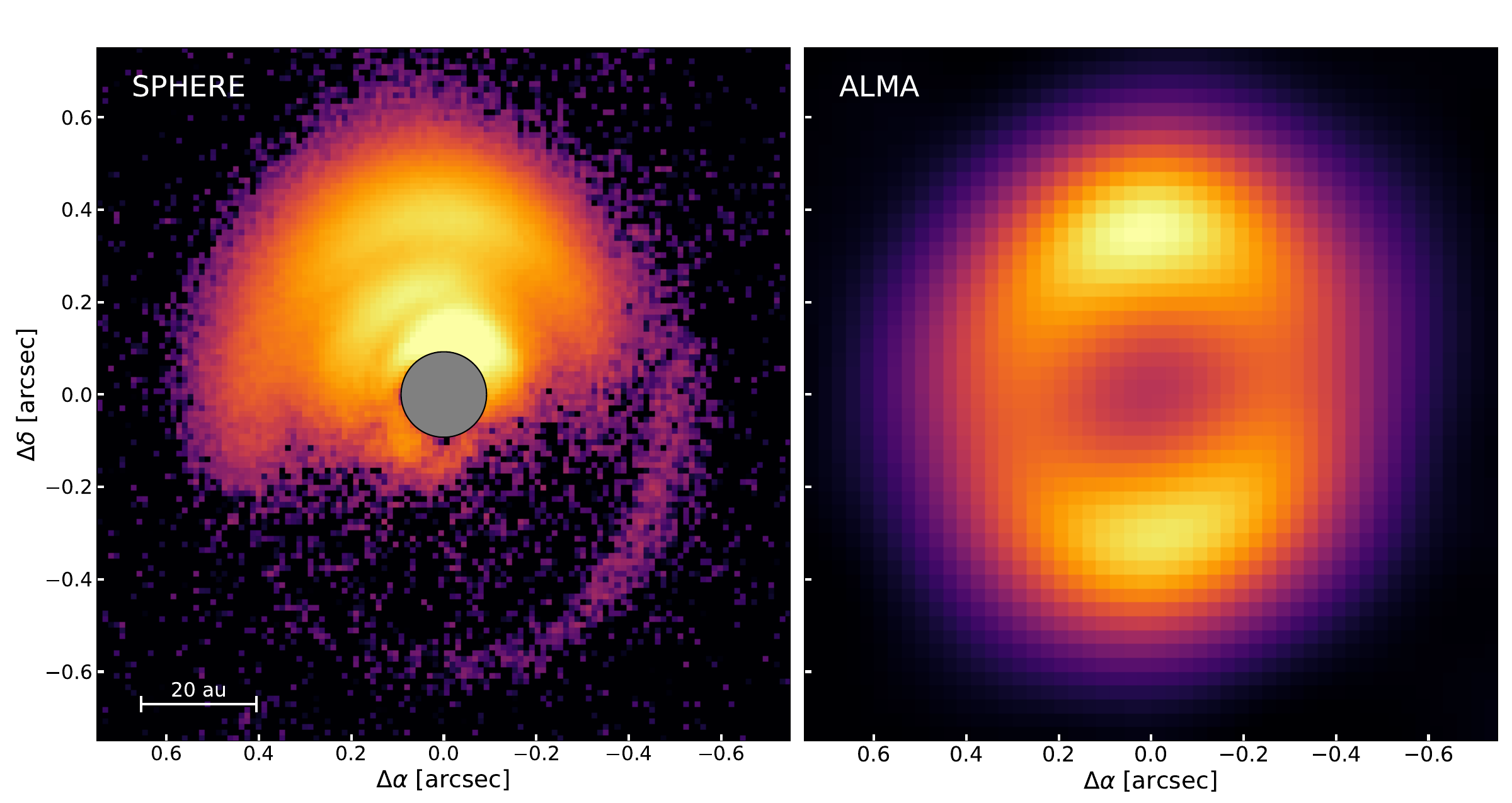}
\caption{SPHERE and ALMA images of the disk around V1098\,Sco showing, respectively, scattered light from the disk surface at 1.6\,$\mu$m (left panel) and thermal emission from larger dust grains at 0.8\,mm (right panel). The SPHERE image is shown on a log scale with a dynamic range of 100 and a coronagraph mask of 92.5 milli-arcseconds. The ALMA image is shown on a linear scale from 0 to the peak value. The resolution of the SPHERE image is about 40 milli-arcseconds and is about ten times larger, $\sim 0\farcs4 \times 0\farcs3$, for the ALMA image.
\label{fig:SPHEREALMA}}
\end{center}
\end{figure*}

\subsection{Disk image comparison}
Images of the disk in scattered light at 1.6\,$\mu$m and thermal continuum at 0.88\,mm are shown in Figure~\ref{fig:SPHEREALMA}.
The SPHERE image shows a large shadow and faint arc of emission to the south, and two annular rings in the north. The ALMA image is strikingly different with a full circle of emission around a large central cavity.
It is apparent that the larger, approximately millimeter-sized dust grains, that produce the ALMA emission are confined to a ring in the outer disk whereas the smaller, approximately micron-sized, grains that scatter infrared light are more broadly distributed but also that there must be some inner disk feature that blocks the starlight from hitting one side of the disk.

Given the very different appearance of the two images, which differ by a factor of $> 500$ in wavelength, we model each dataset independently. We begin by analyzing the interferometer visibilities to determine the properties of the outer disk, then study the kinematics to measure the central stellar mass, and finally perform radiative transfer models to understand the structure of the inner disk and, in particular, how it causes the shadow-arc combination seen in the south.

\section{Geometry of the Outer disk}
\label{sec:outer_disk}
\subsection{Continuum}
\label{sec:galario}
The resolution of the ALMA data is about an order of magnitude lower than that of the SPHERE image but is still sufficient to measure the outer disk properties.
We carry out the analysis in the visibility plane as these are the direct interferometric measurements and avoids additional modeling steps due to the incomplete sampling of the Fourier plane.

\begin{figure}[h!]
\begin{center}
\includegraphics[scale=0.4]{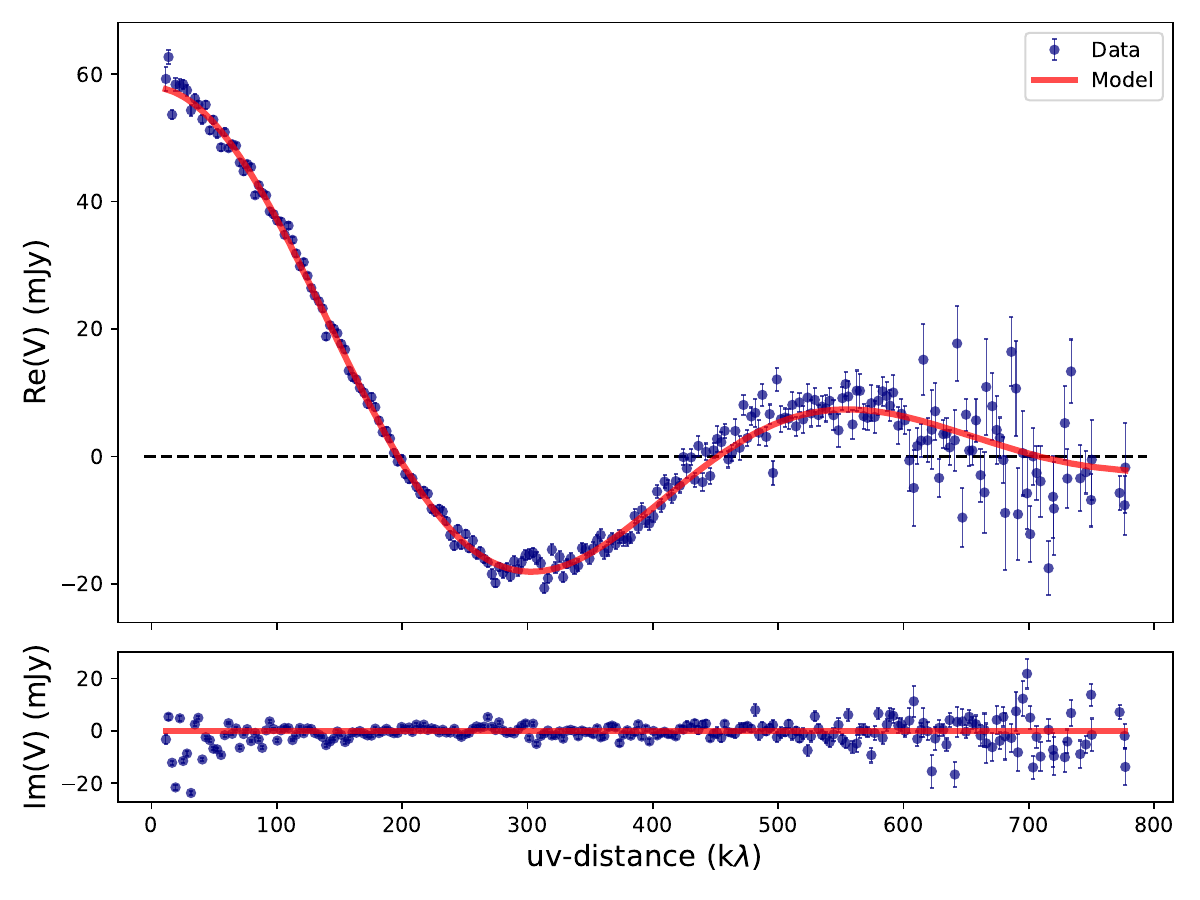}
\caption{Observations and model fit to the real and imaginary parts of the complex ALMA visibilities.
\label{fig:galario}}
\end{center}
\end{figure}

We model the radial intensity profile as an azimuthally symmetric Gaussian ring centered at radius $R_{\rm ring}$ with FWHM $=2.355\sigma_{\rm ring}$,
\begin{equation}
I(R) = I_0\exp\left[-\frac{(R-R_{\rm ring})^2}{2\sigma_{\rm ring}^2}\right].
\end{equation}
Using the {\tt galario} package \citep{galario}, we then fit the real and imaginary parts of the complex visibilities to constrain the ring parameters and disk inclination and position angle. The fit to the data is shown in Figure~\ref{fig:galario} and the inferred parameters are listed in Table~\ref{tab:galario}.

The relatively large detected grains that emit most strongly at these wavelengths are distributed in a wide ring and there is no detectable millimeter emission within $\sim 40$\,au of the star. This is indicative of a strong dust trap, similar to those found in $\sim10$\% of young protoplanetary disks \citep{2018ApJ...854..177V}.

\begin{deluxetable*}{llll}
\tablewidth{0pt}
\tablecaption{ALMA continuum fit\label{tab:galario}}
\tablehead{
\colhead{Parameter} & \colhead{Value} & \colhead{Error} & \colhead{Unit}
}
\startdata
$R_{\rm ring}$ & 63.0 & 0.1 & au \\
FWHM & 29.0 & 0.4 & au \\
incl & 40.1 & 0.25 & deg \\
PA & -2.0 & 0.4 & deg \\
\enddata
\end{deluxetable*}

\subsection{Kinematics}
\label{sec:eddy}
\begin{figure}[h!]
\begin{center}
\includegraphics[scale=0.65]{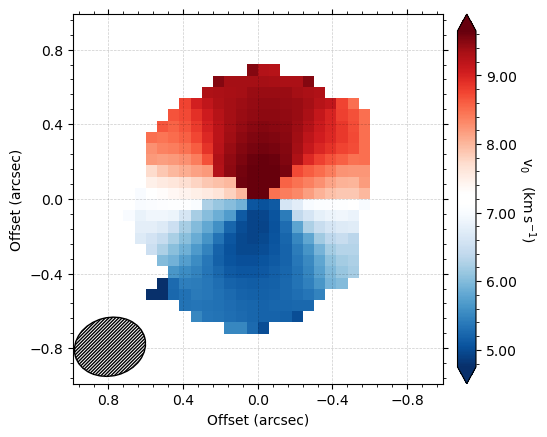}
\caption{Disk rotation map determined the central velocity of the CO 2--1 emission. The $0\farcs38 \times 0\farcs30$ beamsize is shown in the lower left corner.
\label{fig:rotationmap}}
\end{center}
\end{figure}

Although the large dust grains are confined to the outer disk, we detect CO emission within the millimeter continuum cavity. This allows us to map the disk rotation and determine the stellar mass. The first moment of the CO data cube in Figure~\ref{fig:rotationmap} shows a velocity shift of about $\pm 2$\,\kms\ from north to south relative to the systemic velocity.

\begin{figure}[h!]
\begin{center}
\includegraphics[scale=0.3]{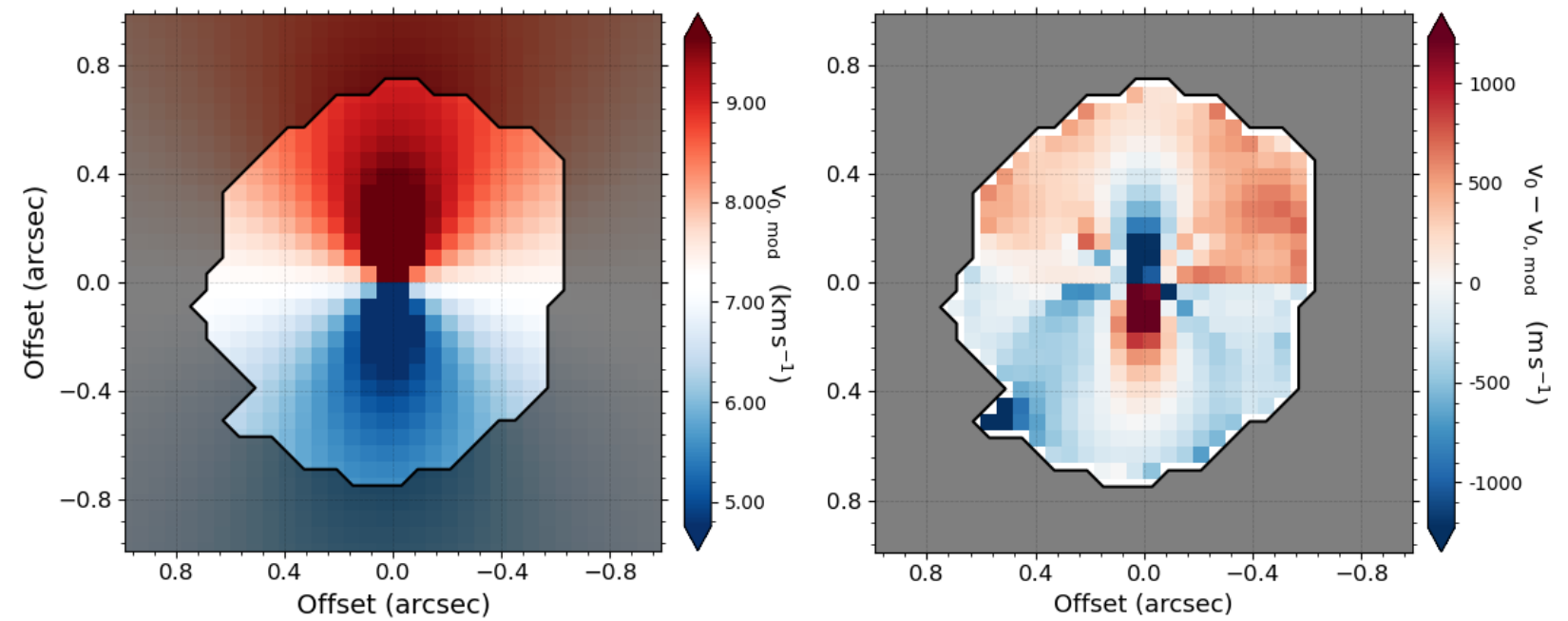}
\caption{Model fit (left panel) and residuals (right panel) to the CO rotation map assuming a flat disk Keplerian profile.
\label{fig:eddy}}
\end{center}
\end{figure}

We use the {\tt eddy} package \citep{eddy} to fit a Keplerian profile to the 2-D velocity field. We fix the inclination equal to $40.1^\circ$, as derived from the galario fit to the continuum visibilities and find a best fit stellar mass of $0.99\,M_\odot$. The systemic velocity is 7.3\,\kms\ and the position angle is $-7^\circ$ (defined as the angle, east of north, of the redshifted side of the major axis). The fit and residuals are shown in Figure~\ref{fig:eddy}.

The derived stellar mass is consistent with the K1 spectral type and 5\,Myr median age of Upper Sco. The position angle is consistent with the {\tt galario} value given the different scale heights of the millimeter continuum and molecular gas emitting regions and the small number of resolution elements across the disk. The formal errors in the {\tt eddy} fit are small (less than 1\%) but the true uncertainties are closer to $\sim 10$\%, due to systematics and additional effects not modeled here including disk pressure and self-gravity \citep{2025ApJ...984L..17L}. The residual map shows that the fit is poor toward the center suggestive of non-Keplerian motions or geometrical effects in this region. The resolution of these data is too poor to examine this further but we surmise that is caused by a tilted inner disk that produces the shadow in the scattered light.

\section{Radiative Transfer Modeling}
\label{sec:radmc}
We model the thermal emission and scattering from the dust using the versatile {\tt radmc3d} code \citep{radmc3d} to fit the ALMA and SPHERE images in turn. This requires a dust density and opacity prescription. We set up the density distribution using a power law in surface density and vertical scale height in cylindrical coordinates, $(R,Z)$, centered on the star,
\begin{equation}
\Sigma_{\rm dust}(R) = \Sigma_1\left(\frac{R}{1\,{\rm au}}\right)^{-\alpha};~~~~R_{\rm in}<R<R_{\rm out},
\end{equation}
\begin{equation}
H(R) = H_1\left(\frac{R}{1\,{\rm au}}\right)^\beta.
\end{equation}
Assuming a vertical isothermal temperature profile, the dust volume density is then given by
\begin{equation}
\rho_{\rm dust}(R,Z) = \Sigma_{\rm dust}(R)\exp\left[-\frac{Z^2}{2H(R)^2}\right].
\end{equation}

We chose this simple truncated power law formulation rather than an exponential taper edge for simplicity. In practice, since the scattered light falls off as the inverse square of radius and the resolution of the ALMA data is low, we are not able to constrain the fall off in surface density at the outer edge of the disk.

We model the dust as consisting of two size populations \citep{2012A&A...539A.148B} with $0.1\,\mu$m and $300\,\mu$m grains and opacities calculated for a composition of amorphous olivine with 50\% Mg and 50\% Fe \citep{1994A&A...292..641J, 1995A&A...300..503D} using the {\tt makeDustOpac} module in {\tt radmc3d}.
Each population follows the same prescription described above but with different parameter sets $\{\Sigma_1, \alpha, R_{\rm in}, R_{\rm out}, H_1, \beta\}$ that we preface with the superscript small or big.

\subsection{Large Grains in the Outer Disk}
The small grains have negligible thermal emission at millimeter wavelengths and the large grains do not efficiently scatter the short wavelength starlight.
We set the mass fraction to be 10\% small grains, 90\% large based on models that match the infrared-millimeter SED of typical disks \citep{2006ApJ...638..314D} though the actual ratio is not critical to this work as we are mainly interested in determining the geometry, rather than the mass or composition of the disk.
We then adjust the density prescription parameters for the large grains to find a representative fit to the ALMA map and then use that to inform the outer disk parameters when carrying out a more detailed fit to the SPHERE map in the following section.

\begin{figure}[h!]
\begin{center}
\includegraphics[scale=0.2]{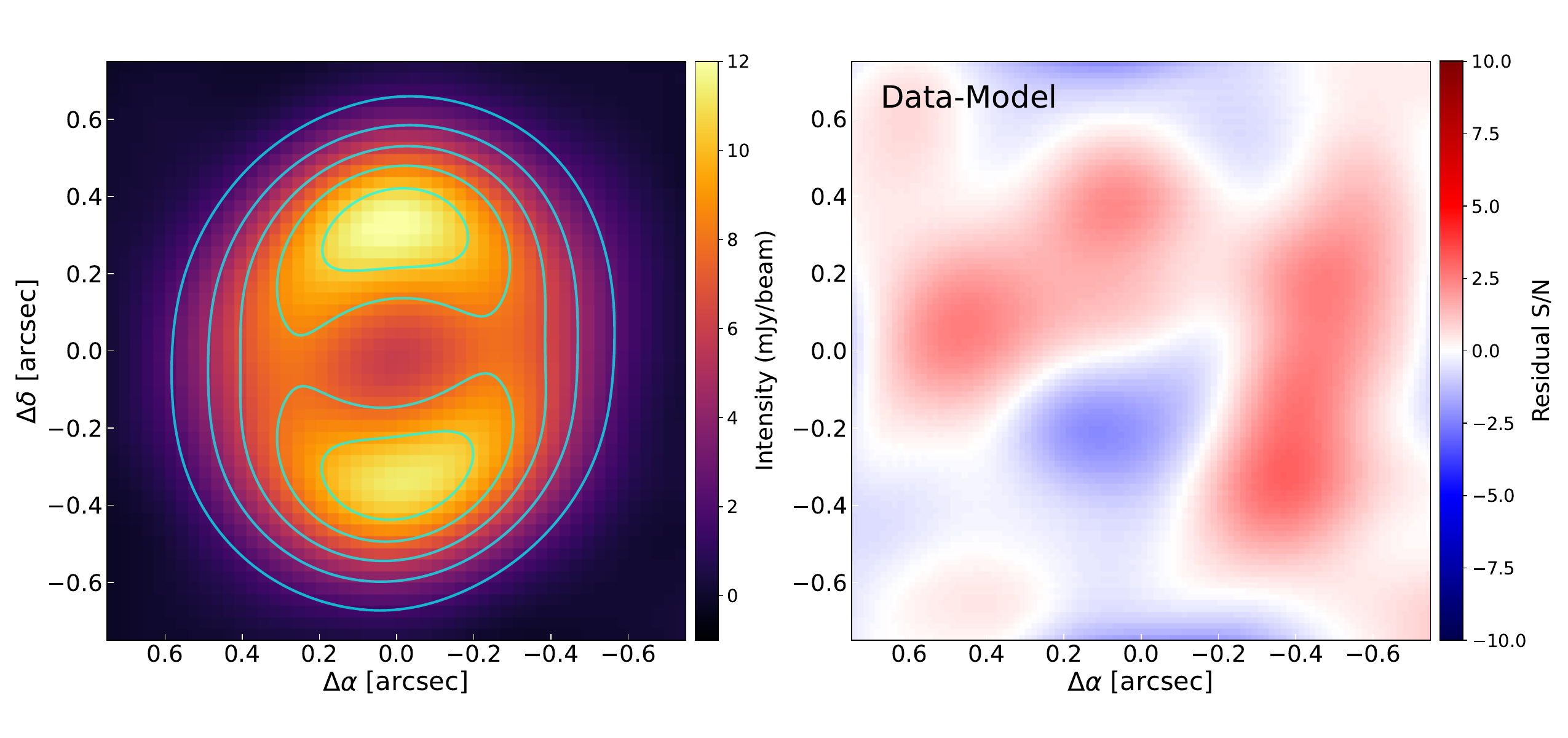}
\caption{Radiative transfer model to the ALMA continuum image. The left panel shows the observed intensities on a linear stretched color scale with the model overlayed in contours from 2 to 12 mJy\,beam$^{-1}$. The right panel plots the residuals in units of the rms noise level in the data.
\label{fig:ALMA_radmc}}
\end{center}
\end{figure}

To model the ALMA continuum image, we fixed the inclination to $40^\circ$ and position angle $-5^\circ$ based on the {\tt galario} and {\tt eddy} fits and ran a grid of models varying $\{\Sigma_1, \alpha, R_{\rm in}, R_{\rm out}\}^{\rm big}$. The quality of each model was assessed by calculating the sum of the squared intensity residuals with the lowest value at $\Sigma_1^{\rm big}=2.2$\,g\,cm$^{-2}, \alpha^{\rm big}=1$ and a radial extent, $R_{\rm in}^{\rm big}\simeq 40$\,au to $R_{\rm out}^{\rm big}\simeq 85$\,au. The total dust mass is $42\,M_\oplus$. As the emission at these wavelengths is optically thin and the disk is seen fairly face-on, our models are insensitive to the vertical scale height parameters, $H_1^{\rm big}$ and $\beta^{\rm big}$, at least at the resolution of the data
A comparison of the data and model are shown in Figure~\ref{fig:ALMA_radmc}.

Given the moderate resolution and dynamic range of the ALMA data, we do not perform a statistically rigorous fit as these constraints on the outer disk are sufficient for the main goal of measuring the properties of the inner disk.

\subsection{Geometry of the Tilted Inner disk}
\label{sec:tilt}
\begin{figure*}
\begin{center}
\includegraphics[width=\textwidth]{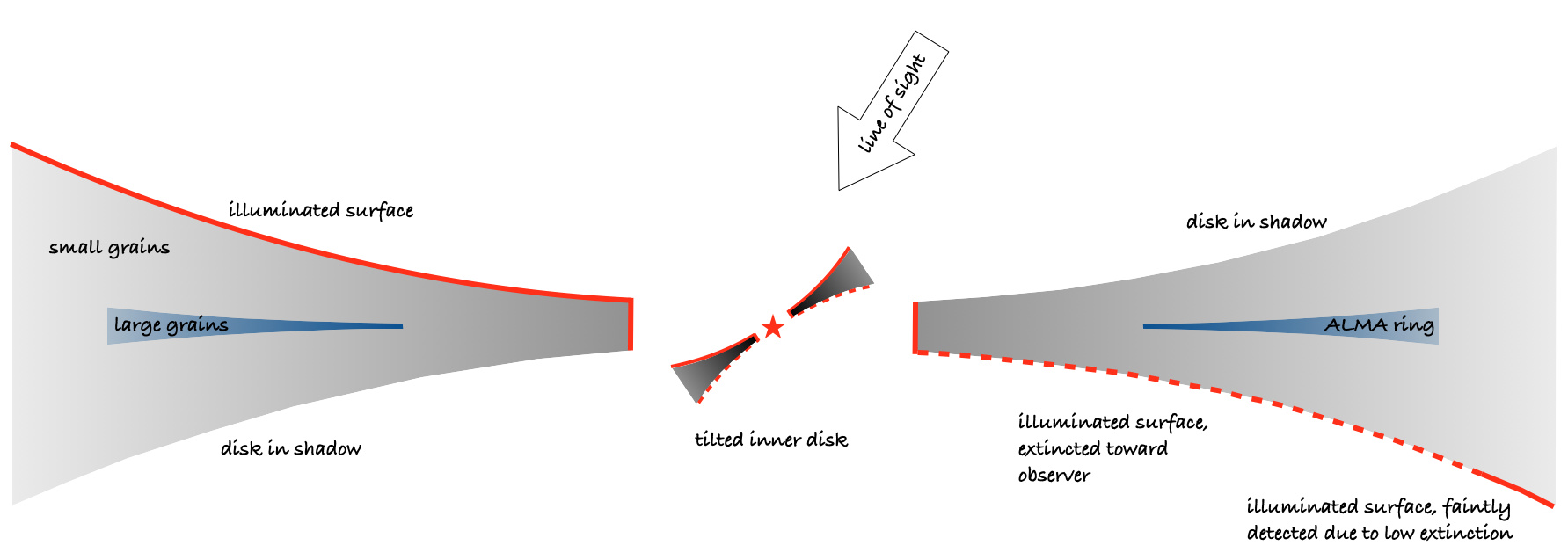}
\caption{Edge-on cutaway of the inner and outer disk geometry. Our viewing angle is from the top right corner such that we see scattered light from the illuminated northern (left) side but not from the shadowed southern (right) side, except for faint emission from the outer parts of the highly extincted backside of the disk. The flattened thick ring of large dust grains seen in the ALMA continuum image is illustrated by the blue region in the midplane.
\label{fig:schematic}}
\end{center}
\end{figure*}
We now proceed to model the shadow seen in the SPHERE scattered light image. Given the symmetry of the ALMA continuum and CO maps and the residuals to the Keplerian fit, we consider a misaligned inner disk.
Disk shadows have been seen in scattered light images of several protoplanetary disks and provide a window into the inner disk often at scales that are hard to image in other ways \citep{2022A&A...658A.183B}.
For V1098\,Sco, the shadow extends across an entire half of the disk similar to HD\,143006 \citep{2018A&A...619A.171B} and HD\,139614 \citep{2020A&A...635A.121M}
and we consider a similar geometric arrangement here, schematically illustrated in Figure~\ref{fig:schematic}.

In this case, we see an additional feature in the form of a faint arc to the south in the otherwise shadowed region. We rule out a scenario where this comes from the outer radii of the front side of the disk, just beyond a shadowed region, as the arc does not extend all the way around the disk nor is it contiguous with the outer parts of the unshadowed, northern disk. Instead, we consider it to arise from the backside of the disk. By symmetry, a tilted inner disk that shadows the southern half of the front side of a flared disk would shadow the northern half of the back side but leave the southern half fully illuminated. As in IM Lup \citep{2018ApJ...863...44A}, we see the extincted outer edge of this back side but in this case only as a partial arc. This fortuitous arrangement provides an extra, critical, constraint on the inner disk geometry.

We fix the parameters for the large grain population based on the ALMA fit shown above and use the same surface density normalization and power law index.
As additional parameters are necessary to describe the inner disk, we first define the overall radial extent. The models are insensitive to the inner edge, $R_{\rm in}^{\rm small}$, as long as it lies within the coronagraph and we fix it to 0.5\,au. The outer radius, $R_{\rm out}^{\rm small}$, simply scales the extent of the emission and we set it to 96\,au ($0\farcs6$) by eye.

To create the shadow, we break the disk at a tilt radius, \Rtilt, and incline and rotate the inner disk at angles \itilt, \PAtilt\ respectively. This requires a change to a 3-D spherical coordinate system $(r, \theta, \phi)$. The tilt radius is constrained to lie within the large grain cavity of 40\,au so that the tilted inner disk is entirely composed of small grains. We note that the SPHERE image shows a bright ring of emission around the coronagraph that extends into the otherwise shadowed southern half, though with a marked asymmetry, which is likely to be either the front side of the tilted inner disk or the inner edge of the outer disk illuminated due to light coming through the dust sublimation hole in the inner disk. The coronagraph hole corresponds to 7.5\,au in radius and we find that our models require a larger \Rtilt\ to match the extent and sharpness of the shadow which implies that we can indeed directly see the inner disk.

\begin{figure*}
\begin{center}
\includegraphics[width=\textwidth]{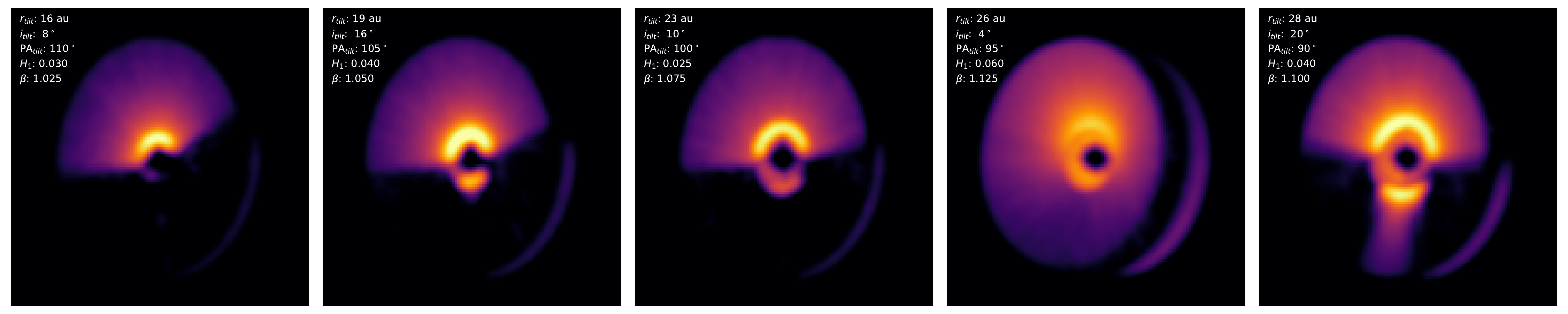}
\caption{A representative set of images showcasing the range of scattered light morphologies in the model grid. The parameter values for each model are shown in the top left corner of each panel. The central region is masked out to mimic the SPHERE coronagraph. Additional plots that show the systematic effects of changing a single parameter one at a time are shown in the Appendix.
\label{fig:representative_images}}
\end{center}
\end{figure*}

The five free parameters\footnote{We drop the suffix ``small'', for readability.}, $\{$\Rtilt, \itilt, \PAtilt, $H_1, \beta\}$, that describe the inner disk geometry and disk vertical structure each affect the scattered light image in distinct ways. A small set of representative models that vary multiple parameters at once is shown in Figure~\ref{fig:representative_images}. The Appendices contain multiple plots where only one parameter is changed at a time to more clearly illustrate their individual affect.

Some general trends are that \Rtilt\ affects the location of the bright ring of emission closer to the star and how clearly it is seen on the southern side. Low \itilt\ produce a blurry shadow boundary, high \itilt\ produce a tongue of scattered light in the shadowed region where starlight escapes through the central hole (we verified that this effect is seen for different $R_{\rm in}$ greater than the dust sublimation radius of $\sim 0.1$\,au and essentially independent of its precise value). \PAtilt\ simply rotates the shadow boundary and arc. The flaring parameters, $H_1$ and $\beta$, affect the illumination on the northern side of the disk, the angle of the shadow boundary, and the brightness of the arc.

We find that we can produce models that meet the ``eyeball-test'' of qualitatively matching the data. To find the best set of parameters that match the observations, we ran a large grid of models stepping through each parameter. The grid range and step size are shown in Table~\ref{tab:model_grid} and amount to 34,560 
models in total.

Whereas we could readily calculate a likelihood for the model fit to the ALMA data based on the squared difference of the intensities, this is much harder to do for the SPHERE image for several reasons. First, the disk shows substructure in the form of a second ring further out from the presumed inner-outer disk boundary. Though interesting in its own right, it would require additional parameters to model that add to the complexity without aiding the solution of what is causing the disk shadow. Second, the scattering is angle dependent but modeling the phase function requires additional parameters for the dust opacities and again this complicates rather than aids the solution. Finally, the faint arc from the backside provides important information but it is about 100 times fainter than the disk from the front side and would therefore be comparably down-weighted in any intensity-based comparison. Moreover, the brightness is a combination of back scattering from the dust surface and extinction by dust in the disk interior so, again, is sensitive to disk properties that are unrelated to our principal goal.

\begin{deluxetable*}{lccc}
\tablewidth{0pt}
\tablecaption{Model grid parameters\label{tab:model_grid}}
\tablehead{
\colhead{Parameter} & \colhead{Range} & \colhead{Step} & \colhead{Unit}
}
\startdata
$R_{\rm tilt}$ & [15, 30] & 1 & au \\
$i_{\rm tilt}$ & [4, 20] & 2 & deg \\
${\rm PA}_{\rm tilt}$ & [90, 110] & 5 & deg \\
$H_1$ & [0.025, 0.060] & 0.005 & au \\
$\beta$ & [1.025, 1.150] & 0.025 & -- \\
\enddata
\end{deluxetable*}

\begin{figure}
\begin{center}
\includegraphics[scale=0.18]{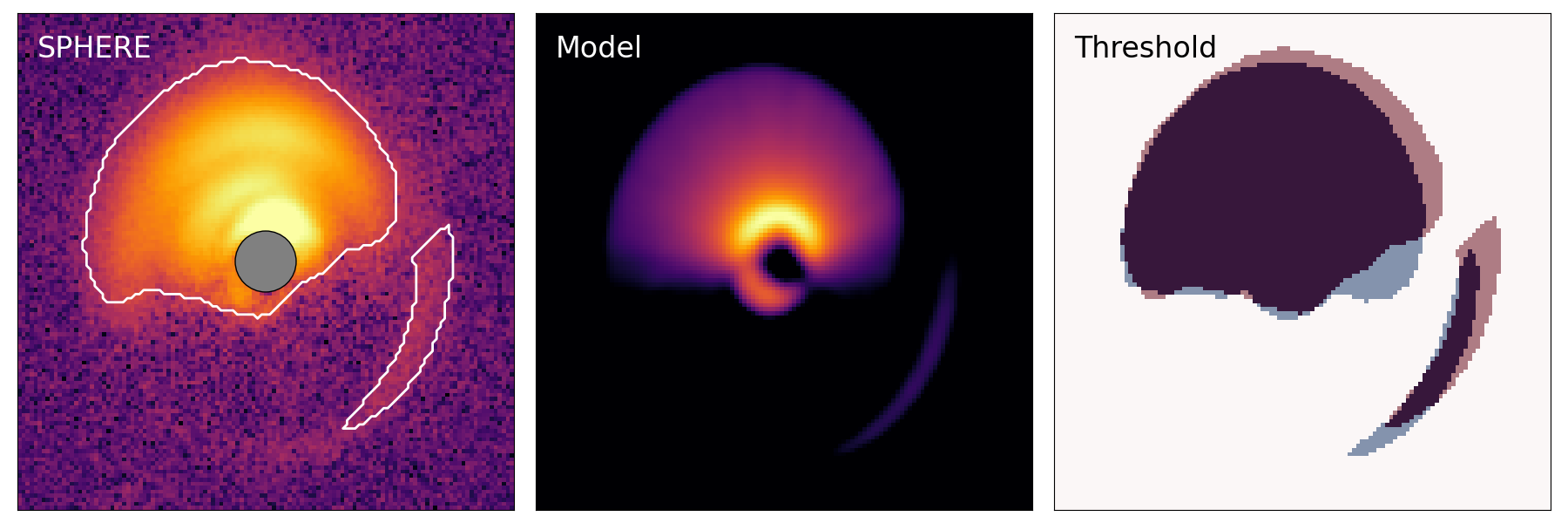}
\caption{Illustration of how the models are compared to the observations. The left panel shows the SPHERE image with borders defined by contouring to highlight the illuminated, northern side of the disk and the faint southern arc. A representative model is shown in the middle panel. An auto-thresholding algorithm then defines the regions of emission. The right panel shows binary images of the emission for the observations (red) and model (blue) which is used to quantify the goodness of fit.
\label{fig:mask}}
\end{center}
\end{figure}

Instead, we focus on matching the scattered light morphology to compare the models to the observations. We define boundaries for the northern disk and southern arc based on contouring the SPHERE image. We then use a local threshold algorithm to define the regions of emission in the model. In this way, we reduce the data and models to binary maps and can readily quantify the goodness of fit by maximizing the likelihood defined by the sum of the squared difference map (equivalent to the sum of the absolute differences). The procedure is graphically illustrated in Figure~\ref{fig:mask}.
Various thresholding algorithms are implemented in the python image processing package {\tt scikit-image} and by experimentation, we settled on the triangle method, where the threshold is automatically determined from the normalized histogram of pixel values \citep{doi:10.1177/25.7.70454}.

Using this technique, we are able to quantify the model fits to the SPHERE image and constrain the range of parameters that provide good fits. This shows that the model design of a tilted inner disk works well and that we can constrain its properties. However, to determine the best fit parameter uncertainties and understand their inter-dependences requires a new approach since the radiative transfer model is computationally expensive. This motivates a machine learning approach.

\subsubsection{Model Image Interpolation Using an Artificial Neural Network}
\label{sec:neural_network}

\begin{figure*}
\begin{center}
\includegraphics[width=\textwidth]{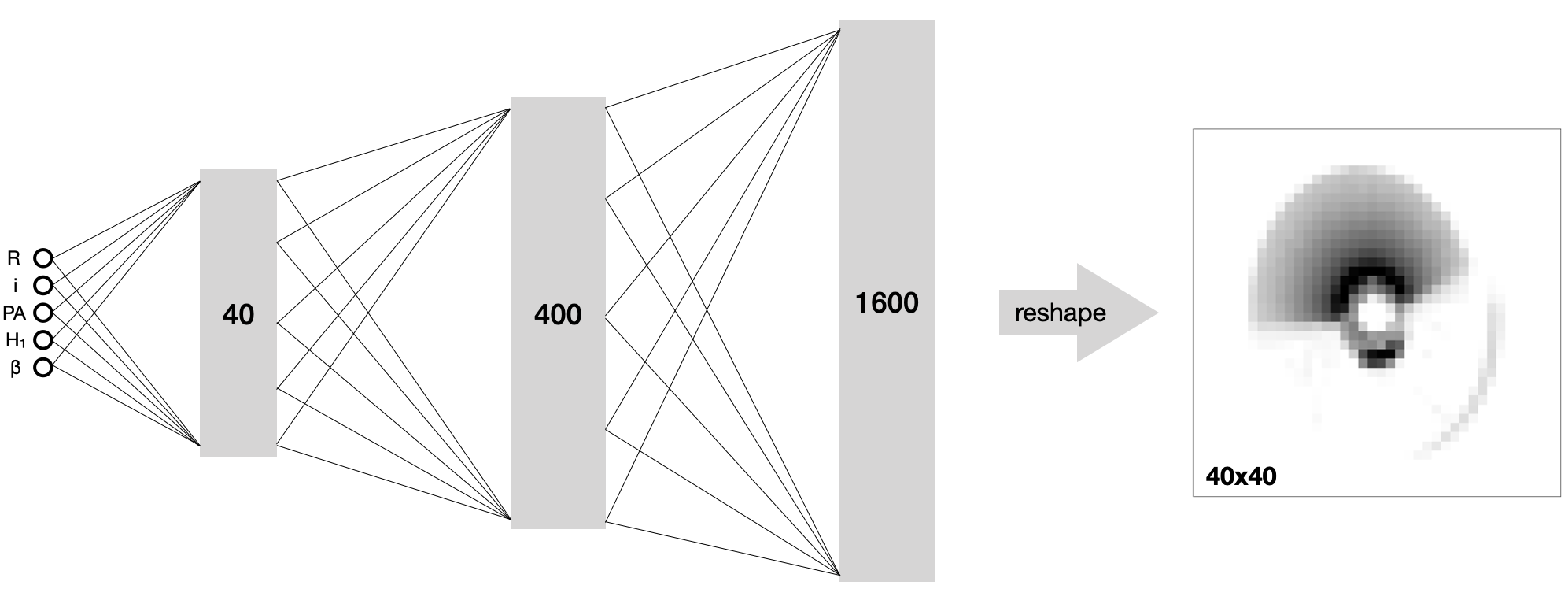}
\caption{The architecture of the neural network used to produce images from the five input parameters of the model. There are three layers in all, represented by the grey rectangles and labeled by the number of output nodes. The inter-connections between them are schematically shown (in practice each output node of one layer connects to every input  node of the next layer). The final layer produces a vector with 1600 elements which we reshape into a $40\times 40$ pixel image.
\label{fig:neural_network}}
\end{center}
\end{figure*}

Each {\tt radmc3d} image is produced by propagating photons through the disk in random directions and modeling the scattering with dust particles by sampling from the allowed range of quantum outcomes. To reduce the inherent Poisson noise in this method requires a large number of photons and with a correspondingly high computational demand. For the initial parameter exploration, we use $10^6$ photon packets and each model takes almost two minutes to run. Even with multiple cores, the grid of $\sim 35,000$ models takes many days to produce. To carry out an MCMC parameter search would likely require at least this many and perhaps as much as an order of magnitude more models to achieve statistical convergence. Furthermore any tweaks to the model design would require a restart. In short, this approach is extremely time intensive and consequently a Bayesian approach to radiative transfer modeling is rarely used \citep{2017ApJ...851...45S, 2024A&A...681A..19C}.

An alternative approach is to interpolate model images at intermediate parameter values between the grid points. \citet{2016ApJ...830...32W} carried out an ad hoc formulaic approach to a set of simulated ALMA CO data cubes but more sophisticated techniques are now available that can better handle the complexities of multi-parameter image production and, equally importantly, test and verify the outcome \citep[Simulation-Based Inference:][]{2020PNAS..11730055C}. Here, we train an artificial neural network on radiative transfer images and use it to produce new models at any set of parameter values.
Our approach is similar to \citet{2020A&A...642A.171R, 2023A&A...672A..30K} and \citet{2024A&A...685A..65R} who used neural networks trained on SED models and hydrodynamical simulations respectively.

The network architecture is diagrammed in Figure~\ref{fig:neural_network}. The input is a vector composed of the five model parameters, $\{$\Rtilt, \itilt, \PAtilt, $H_1, \beta\}$, and the output is a vector with 1600 elements that we reshape into a $40\times 40$ image.
The input and output layers are connected through two hidden layers with increasing numbers of neurons to enable the scaling from 5 parameters to the 1600 pixel values. Each neuron in a layer is related to each neuron in the preceding layer by a multiplication factor (weight), an offset (bias), and a non-linear activation function (in this case, ``leaky ReLu'' filters).
The network is trained (i.e., weights and biases are iteratively determined) by comparing the outputs (predictions) to an image dataset (targets) produced with known input parameters. The comparison is quantified via a loss function that calculates the mean squared error between the network predictions and training targets.

Rather than train the network on the existing grid of models, we sampled the parameter space over the same range in Table~\ref{tab:model_grid} using a Sobol sequence as in \citet{2023A&A...672A..30K}. This ensures that the parameter set in each model is different from every other model which is both efficient and mitigates possible overlearning errors from duplicate parameter subsets. Indeed, we found that we could achieve a substantially lower space filling discrepancy than the large grid with just a few hundred Sobol samples. We took advantage of this to increase the signal-to-noise ratio of each radiative transfer model by running {\tt radmc3d} with $5\times 10^6$ photon packets, fives times more than the grid models, and produced a training set of 1024 images.

We found that the loss function decreased rapidly with a few hundred iterative epochs but then more slowly after about a thousand epochs. However, there were occasional spikes that we found to be due to amplification of noisy pixels in the radiative transfer models toward the southern (shadowed) side of the disk. The number of these spikes in the loss function ticked up slightly beyond 2000 iterations. We therefore used the network weights and biases as determined from 2000 epochs for our model fitting.

\begin{figure}
\begin{center}
\includegraphics[scale=0.42]{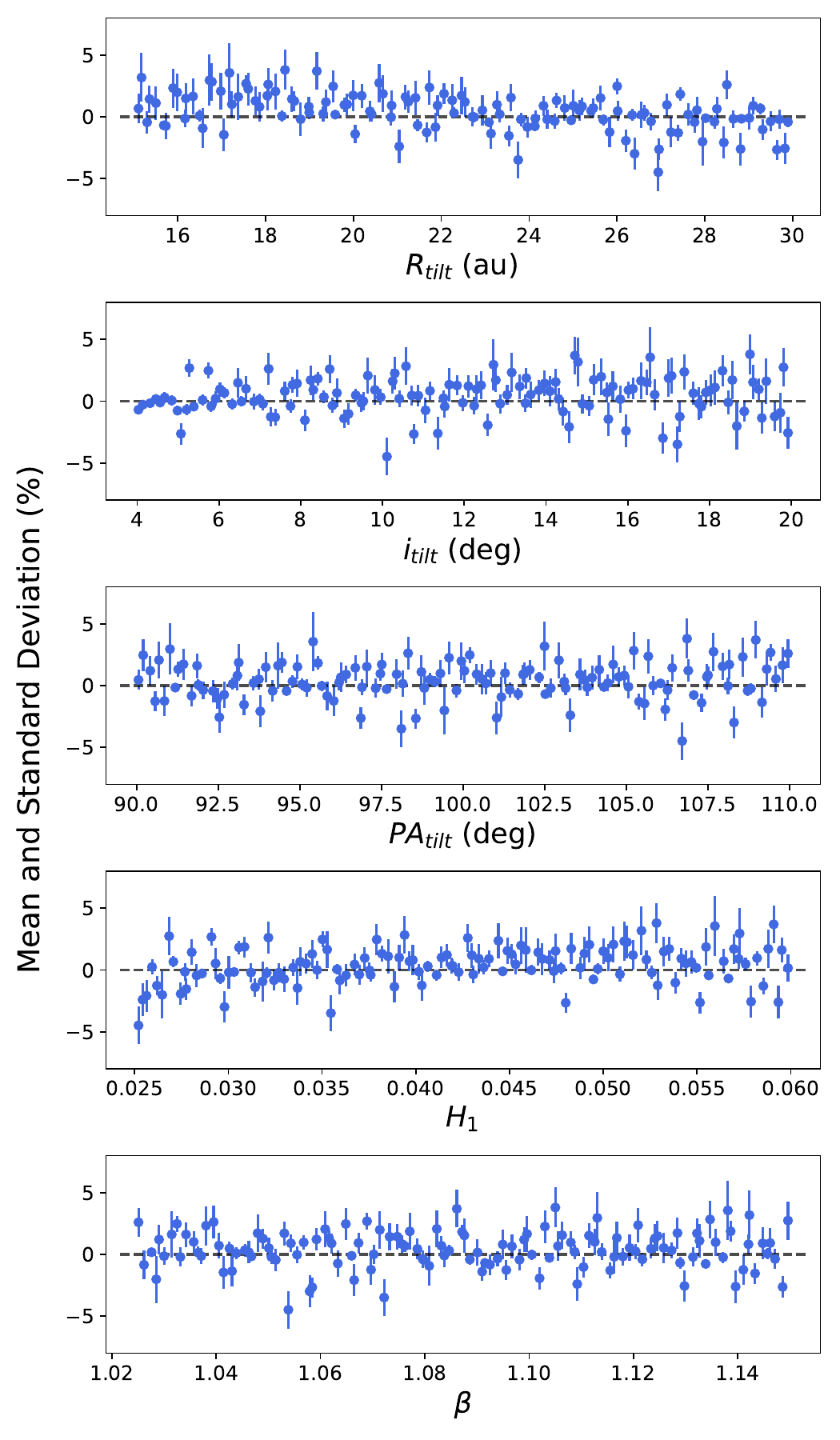}
\caption{Validation test of the interpolation procedure. The mean difference between the images produced by radiative transfer and the neural network is shown for 128 models at random parameter values within the grid boundaries.
\label{fig:validation}}
\end{center}
\end{figure}


\begin{figure*}
\begin{center}
\includegraphics[width=\textwidth]{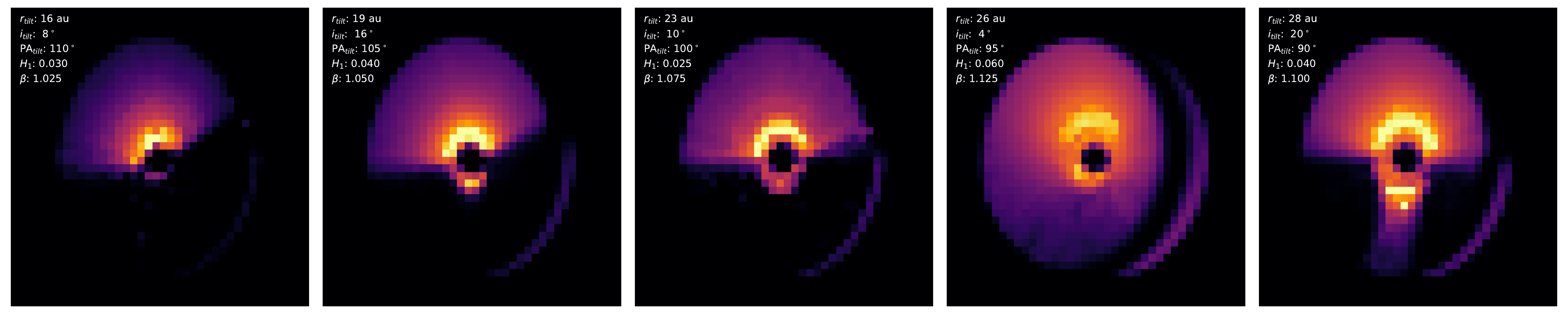}
\caption{A set of images produced by the neural network for the same parameter sets used in Figure~\ref{fig:representative_images} that showcase the range of disk morphologies in the large model grid. The size of these images is $40\times 40$ pixels, three times lower than the {\tt radmc3d} images, but the agreement is otherwise very good.
\label{fig:validation_images}}
\end{center}
\end{figure*}

To demonstrate the network performance, we carried out a validation test by comparing an independent set of 128 radiative transfer models with neural network images from a different Sobol sequence. The results of this validation procedure are shown in Figures~\ref{fig:validation}.
Here the mean and standard deviation are calculated over the $40\times 40$ pixels in each image which is about a factor of two times greater than the area of the physical disk projection. In a few cases, one or a small group of ``hot'' pixels caused by the aforementioned amplification of noise in the southern disk were clipped in the calculation. Higher signal-to-noise radiative transfer models might be required for training neural networks for detailed intensity modeling across a high dynamic range but it is mitigated here by the auto-thresholding algorithm illustrated in Figure~\ref{fig:mask} that we use to fit the data.

The accuracy of an extrapolation from 5 inputs to a 1600-element vector seems at first surprising but there is a high correlation between neighboring pixels due to the smoothness of the physical disk structure.
As a direct visual comparison, Figure~\ref{fig:validation_images}, plots the neural network predictions for the representative set of grid images shown in Figure~\ref{fig:representative_images}. Allowing for the lower resolution of the neural network output, the similarity with the {\tt radmc3d} images across the full range of disk morphologies found in the grid is quite striking but an example of a ``hot'' pixel is seen in the fifth panel where the network is not as well trained for the relatively rare cases where there is a narrow region of scattered light in the southern disk.

We experimented with the network design by changing the number of nodes in each layer, the number of layers, and the activation function.
We found that additional layers produced similarly low loss functions but with a slightly higher proportion of (still occasional) spikes and no significant differences in the validation tests. We therefore settled on the relatively simple architecture in Figure~\ref{fig:neural_network}.

Modern python packages such as {\tt PyTorch} (used here) and {\tt TensorFlow} provide a straightforward setup of artificial neural networks, opening up a new and efficient pathway to fit radiative transfer models to observations. The computational time required to produce high quality images decreases from minutes to milliseconds. This then allows a Bayesian approach to parameter estimation.

\subsubsection{MCMC model fitting}
\label{sec:mcmc}
With the ability to quickly produce model images for any set of physical parameters and using image processing techniques that identify the areas of emission and allow quantitative comparisons with the data, we can search for the best fit through an MCMC analysis. We use the {\tt emcee} package \citep{emcee}, uniformly sampling across the parameter extents of our model grid in Table~\ref{tab:model_grid}, with 20 walkers.

\begin{figure}[h!]
\begin{center}
\includegraphics[scale=0.28]{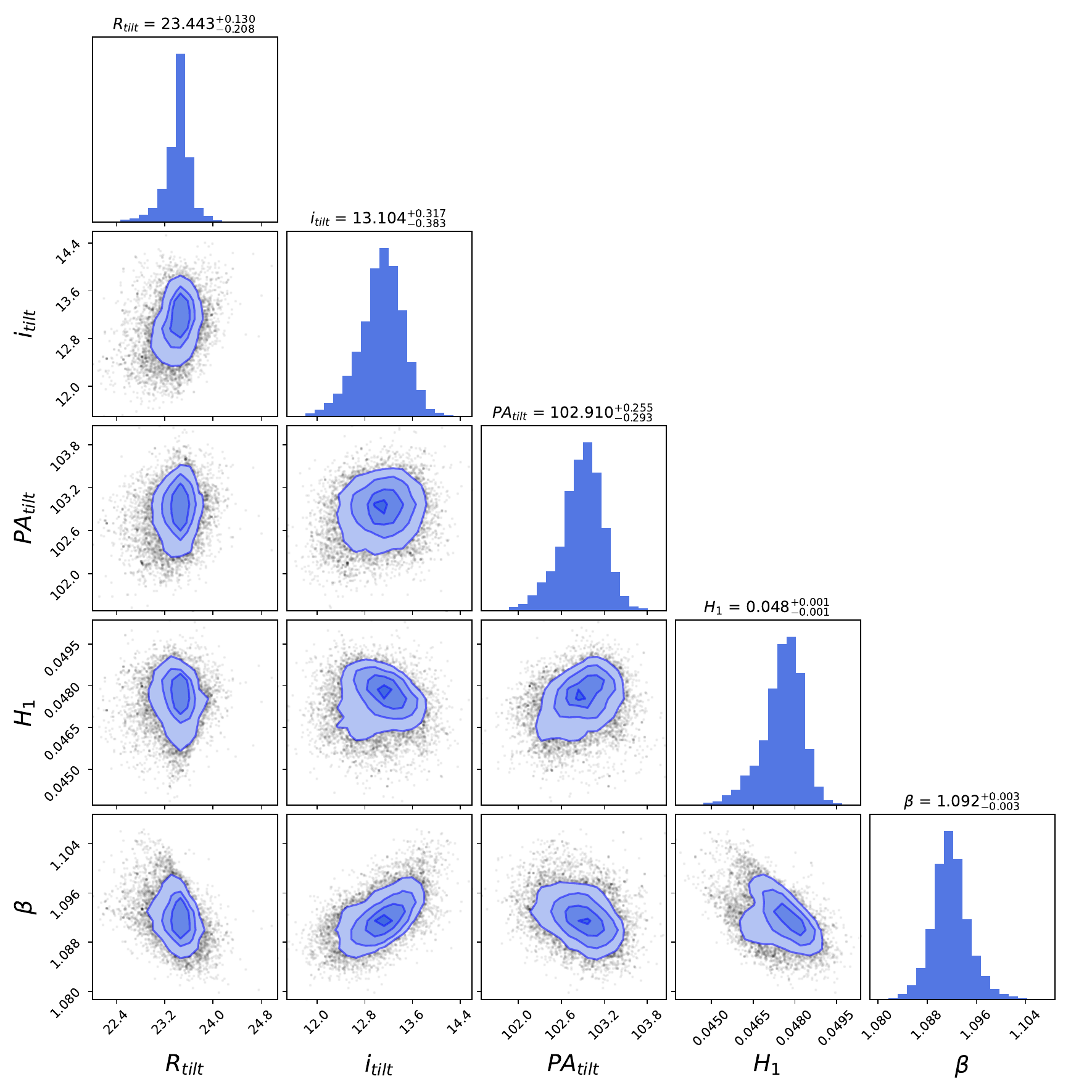}
\end{center}
\caption{Corner plot of the parameter distributions from the MCMC fitting procedure. The histograms at the top of each row show the distributions of a single parameter for all sets of other parameter values. The contour plots show how the fits for each pair of parameters depend on each other.
\label{fig:corner}}
\end{figure}

The MCMC chains converged rapidly and we show results for 10,000 steps after discarding the first 2,000 from the burn-in phase as a corner plot in Figure~\ref{fig:corner}. The contour plots show a correlation between the scale height and flaring index, and a slight dependence with the position angle of the inner disk tilt. The histograms of individual parameters are nearly symmetric and we use the median and maximum of the differences with the 16\% and 84\% percentiles to provide an approximate mean and standard deviation for each parameter in Table~\ref{tab:mcmc}.
The precision with which each parameter is determined belies the error in matching the observations which is mainly due to the validity of the model description rather than the statistical uncertainties in the fit.

\begin{deluxetable*}{llll}
\tablewidth{0pt}
\tablecaption{Inner disk fit\label{tab:mcmc}}
\tablehead{
\colhead{Parameter} & \colhead{Value} & \colhead{Error} & \colhead{Unit}
}
\startdata
\Rtilt  & 23.4 & 0.2 & au \\
\itilt  & 13.1 & 0.4 & deg \\
\PAtilt & 102.9 & 0.3 & deg \\
$H_1$   & 0.048 & 0.001 & au \\
$\beta$ & 1.092 & 0.003 & -- \\
\enddata
\end{deluxetable*}

\begin{figure*}
\begin{center}
\includegraphics[width=\textwidth]{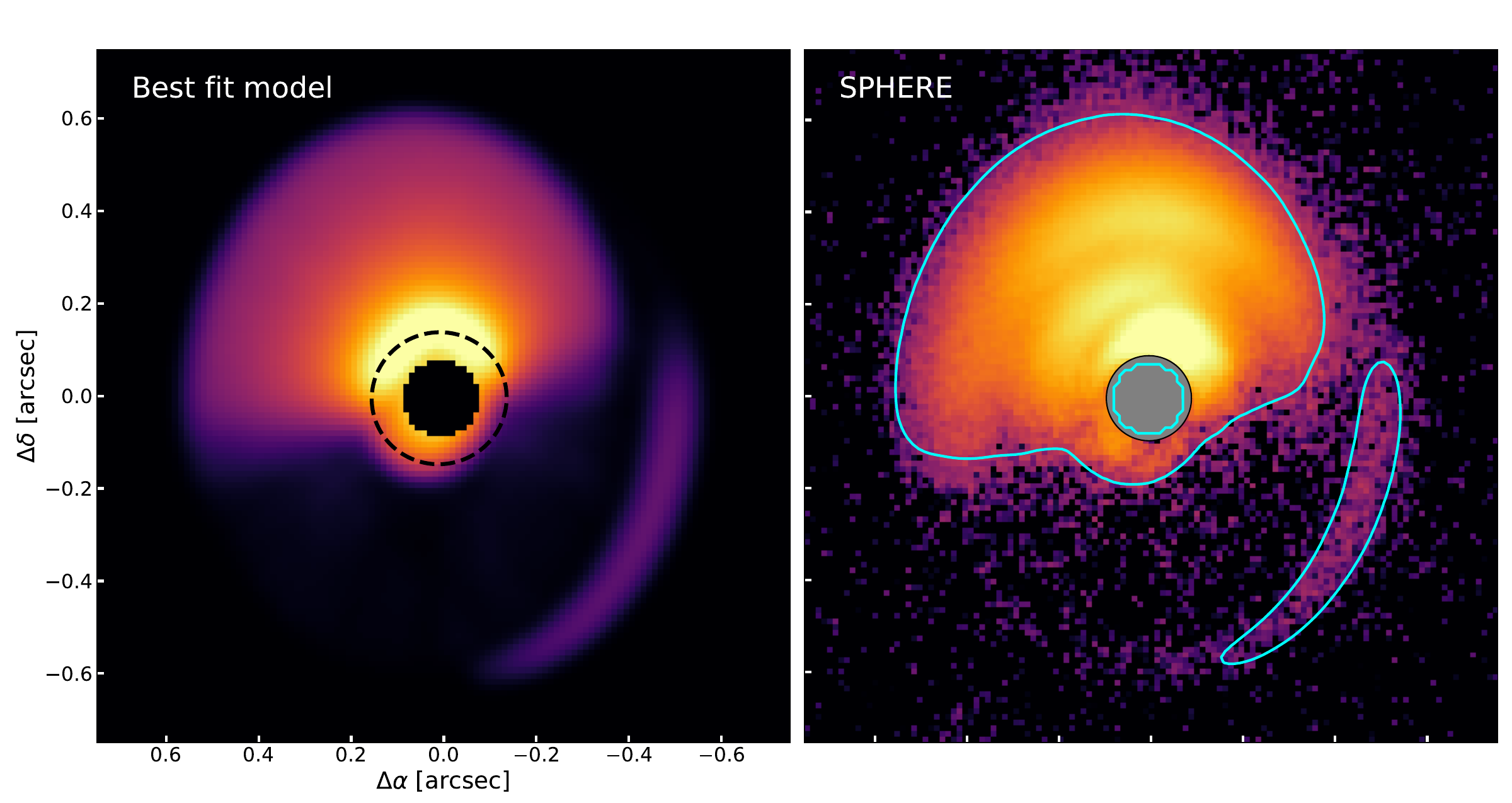}
\end{center}
\caption{Model image for the maximum likelihood model from the MCMC fit (left panel). The dashed ellipse in the model image shows the break radius between the inner and outer disks at 20\,au. The SPHERE image is overlayed with a low level contour of the model to show the level of agreement between the extent of the emitting regions.
\label{fig:bestfit}}
\end{figure*}

Figure~\ref{fig:bestfit} shows the maximum likelihood model in comparison with the SPHERE image. The parameters of this model, $\Rtilt=23.4\,{\rm au}, \itilt=13.0^\circ, \PAtilt=103.0^\circ, H_1=0.048\,{\rm au}, \beta=1.090$ are very similar to the median values from the marginalized histograms in Table~\ref{tab:mcmc}.
With the exception of the disk substructure and scattering phase angle dependence, neither of which we attempted to model, the agreement is very good showing that the five parameters for disk size, flaring, and orientation, are sufficient to capture the essential features in the image and allow us to determine the geometric properties of the inner disk.

\section{Discussion}
\label{sec:discussion}
We have carried out a step-by-step modeling procedure of ALMA and SPHERE observations of V1098\,Sco to explain the remarkable shadow feature that obscures the southern half of the disk. The ALMA continuum observations show a ring centered at a radius of 63\,au around a central cavity depleted of large dust grains within about 40\,au radius. The CO map constrains the stellar mass to about $1\,M_\odot$ but shows residuals near the center that we attribute to a tilted inner disk. Using these results, we fix the inclination and position angle for the outer disk to $i_{\rm outer}=40^\circ$ and ${\rm PA}_{\rm outer}=-5^\circ$ respectively, and were then able to match the morphology scattered light emission with a tilted and misaligned inner disk.

The inferred size of the inner disk has a radius of 23\,au which is large enough that the inner edge of the outer disk lies beyond the SPHERE coronagraph (7.5\,au radius) and appears as a bright ring in the north and a fainter ring in the south, but also lies comfortably within the central hole seen in the ALMA image. The inner disk is inclined by about $13^\circ$ which is more face on than the outer disk but is most different in position angle with a value of  $103^\circ$ for the angle of the major axis relative to North (with a $180^\circ$ geometric degeneracy).
This strong inner disk misalignment explains why the shadow boundary runs from East-West almost orthogonal to the near North-South line of the outer disk major axis. If the inner disk were tilted but had the same position angle as the outer disk, the shadow boundary would be parallel to the major axis.

\begin{figure}
\begin{center}
\includegraphics[scale=0.2]{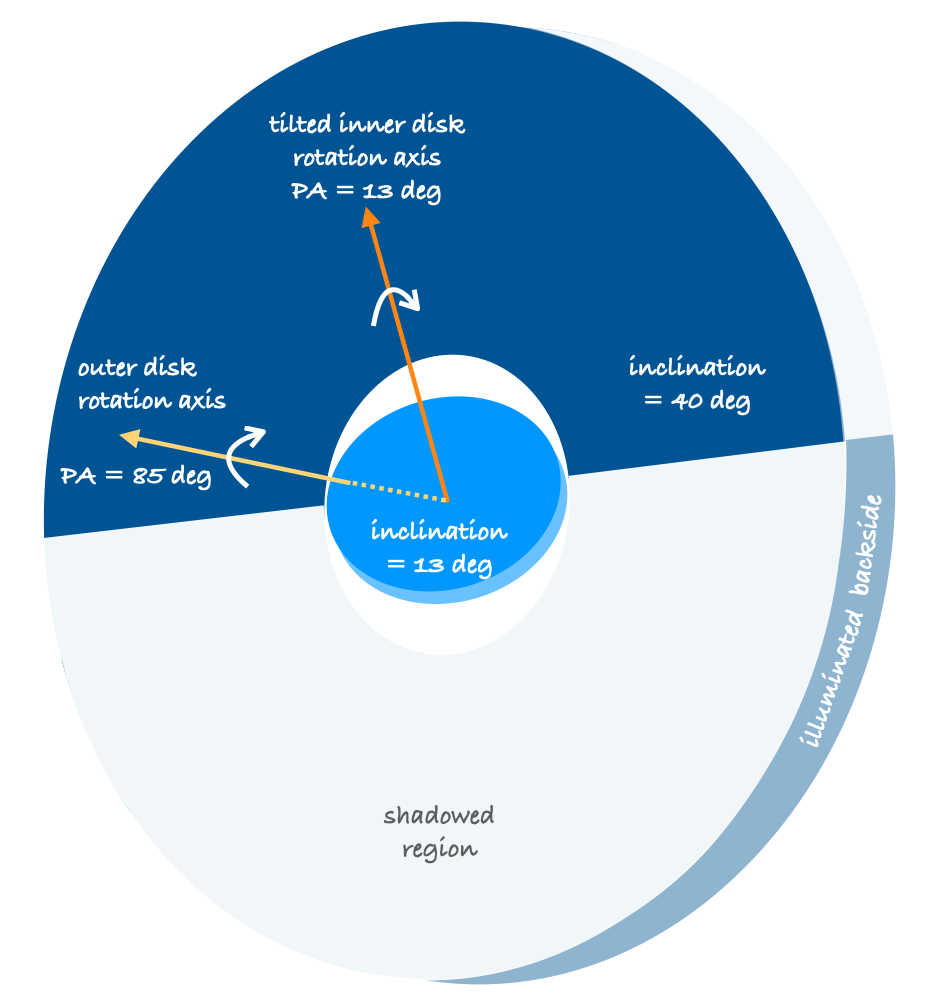}
\end{center}
\caption{Geometry of the disk system inferred from the ALMA and SPHERE modeling. The outer disk is moderately inclined with the major axis running approximately North-South. The inner disk is more face-on and almost orthogonally oriented such that the shadow boundary runs East-West. The inferred misalignment between the rotation axes of the inner and outer disk are shown by the orange and yellow arrows respectively.
\label{fig:disk_rotation_vectors}}
\end{figure}

The more face-on orientation of the inner disk implies smaller projected rotation velocities which explains why the Keplerian fit residuals are opposite to the sense of the overall disk rotation in Figure~\ref{fig:eddy}. Under the reasonable assumption that the inner and outer disk rotate in the same direction, the position angle for the redshifted side of the inner disk is then $-77^\circ$ using the same definition as that used for the outer disk. We then conclude that the difference in the projected angles of the inner and outer disk rotation axes is $\Delta{\rm PA}=72^\circ$, as illustrated in Figure~\ref{fig:disk_rotation_vectors}.
From \citet{2017A&A...604L..10M}, we calculate the misalignment angle between the outer disk and tilted inner disk to be
\begin{equation}
\begin{split}
\mu = \cos^{-1}&[\sin(i_{\rm outer})\sin(\itilt)\cos(\Delta{\rm PA})\\
               &+\cos(i_{\rm outer})\cos(\itilt)] = 38^\circ.
\end{split}
\end{equation}

A misalignment between the inner and outer disk rotation axes requires a gravitational torque from a second source in addition to the central star.
Binary stars can warp a circumstellar or circumbinary disk \citep{2012Natur.491..418B,2013MNRAS.433.2142F} and, for the latter case, a small inner disk around one of the stars can cast shadows on the outer disk \citep{2018MNRAS.477.1270P}. Simulations show that a range of scattered light morphologies can be produced including large shadows that closely resemble our SPHERE observations \citep{2018MNRAS.473.4459F}.

There are no clear signatures of a binary companion.
V1098 Sco lies within a fairly crowded star field and there are several faint optical sources within an arcminute but there are no Gaia sources at the same distance. The large size of the outer disk argues against a companion within several hundred au ($\simlt 2''$).
The Gaia DR3 Renormalized Unit Weight Error (RUWE) of 1.648 is slightly higher than typical values for single stars \citep{2024A&A...688A...1C} but this could be attributed to the presence of scattered light from the disk \citep{2022RNAAS...6...18F}.
The CO velocities analyzed in \S\ref{sec:eddy} show that the central mass is $1\,M_\odot$ which is consistent with the K1 spectral type and age of $\sim 5$\,Myr of the star and rules out a close equal mass binary.

\begin{figure}
\begin{center}
\includegraphics[scale=0.4]{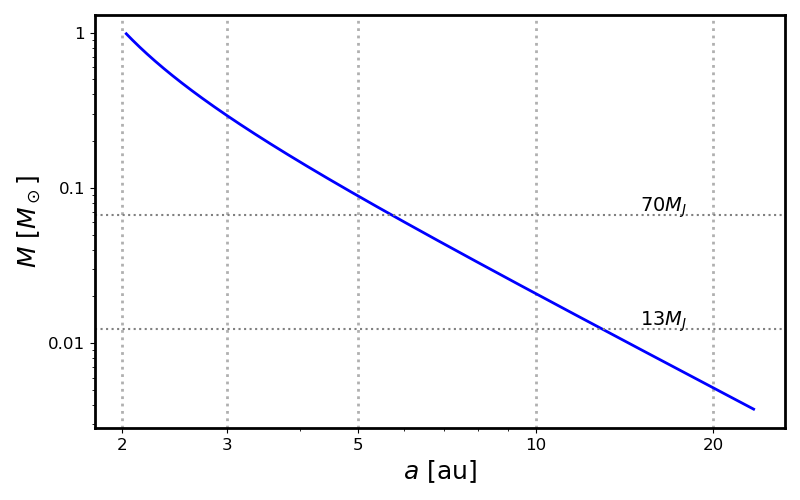}
\end{center}
\caption{The relation between the mass and orbital radius of a companion that would break the disk at 23\,au. The horizontal dotted lines mark the substellar and planet boundaries at 70 and 13 Jupiter masses respectively.
\label{fig:companion_constraint}}
\end{figure}

Nevertheless, the observations allow for a low mass close companion that is too faint to change the spectral type and either lies behind the coronagraph, is enshrouded in dust, or otherwise undetectable in the SPHERE image.
\citet{2013MNRAS.434.1946N} model the evolution of a misaligned gas disk around a binary and show that the break radius between the inner and outer disk depends on the binary separation and mass ratio. Although they focused on disks around supermassive black holes, they also provide a formula for a pressure-dominated disk in an Appendix which is applicable to protoplanetary disks. In our notation, this translates to a constraint on the putative companion orbital radius, $a$, and mass, $M$,
\begin{equation}
a \left(\frac{M}{1\,M_\odot+M}\right)^{1/2} \simgt \Rtilt \left(0.75 \sin2\mu\;\frac{R}{H}\right)^{-1/2},
\end{equation}
where we have substituted the mass of the primary from \S\ref{sec:eddy}.
Our modeling indicates that the aspect ratio at the edge of the inner disk is small, $H/R\simeq 2.7\times 10^{-3}$ and, together with our measurement of the misalignment angle, $\mu=38^\circ$, the right hand side equates to 1.4\,au. We plot the relation between companion mass and orbital radius in Figure~\ref{fig:companion_constraint}.
This holds as long as we are in the wave-like regime of warp propagation which requires a low viscosity, $\alpha < H/R$.

Once a companion opens up a gap, the disk breaks into an inner and outer region which then independently dynamically evolve \citep{2018MNRAS.481...20N}.
A non-zero planet obliquity can then drive a tilt mismatch between the inner and outer disks on Myr timescales which then precess at different rates and become misaligned \citep{2017MNRAS.469.2834O, 2019MNRAS.483.4221Z}.
Future higher resolution studies of the CO kinematics can provide stronger constraints on the mass and location of any perturbing object, or perhaps lead to consideration of multiple objects.

In general, shadowed disks are interesting sources for studying disk physics and hunting for protoplanets. They systematically have high NIR excesses indicative of large inner disks \citep{2018A&A...620A..94G} and ALMA imaging often shows an inner cavity and outer continuum ring which can be explained by a super-Jupiter mass object that prevents millimeter-sized grains from drifting inwards \citep[e.g.,][]{2017ApJ...850...52P, 2019A&A...624A...7V}.
Detailed multi-wavelength radiative transfer models can help us better quantify the properties of these complex sources and guide searches for young planets.

\section{Summary}
\label{sec:summary}
We have imaged the scattered light from the disk around the T Tauri star V1098\,Sco and found a large shadow that covers its entire southern half. Using ALMA archival data, radiative transfer models, and machine learning to speed up image production, we show that the morphology of the scattered light emission can be explained by a disk that is broken into inner and outer regions which are misaligned in both inclination and position angle. The main results of this work are as follows:

\begin{itemize}
\item{The ALMA continuum image shows a central cavity with radius $\sim 40$\,au that is devoid of large (approximately millimeter-sized) dust grains. The CO velocity map constrains the mass of the central star to be $1\,M_\odot$. We use these ALMA observations to fix the inclination and position angle of the outer disk;}
\item{We produce a grid of radiative transfer models and show that a tilted inner disk of small (approximately micron-sized) dust grains can qualitatively reproduce the large shadow and faint arc features in the SPHERE scattered light image of the disk;}
\item{We use an auto-thresholding technique borrowed from text recognition algorithms to identify the salient features in the model images and thereby quantitatively compare with the observations without the extraneous details of disk substructure and scattering phase function;}
\item{We designed a neural network to efficiently produce model images for arbitrary parameter values in between the discrete model grid points. The tremendous increase in speed allows forward modeling of the SPHERE image;}
\item{We constrain the size, inclination, and position angle of the inner disk. It's radius of 23\,au is about half that of the ALMA cavity size. The inner disk is more face-on and twisted such that it's rotation axis is offset by $38^\circ$ from that of the outer disk.}
\end{itemize}

The scattering of starlight from a circumstellar disk depends not only on the spatial distribution of the dust but also its optical properties. The former is more meaningful for understanding planet formation. We have shown that simple image processing techniques that delineate morhpology, coupled with machine learning tools that speed up radiative transfer modeling, provide a Bayesian pathway to quantitative modeling of disk observations with SPHERE. We hope that the methodology described here proves useful for the analysis of other data in the future.

\begin{acknowledgments}
We first acknowledge the referee for an insightful report and the suggestion to use Sobol sequences to sample the parameter space when training the neural network.
We thank Kees Dullemond for providing code to warp the inner disk that we used in the radmc3d modeling.
Thanks also to Richard Nelson and Jaehan Bae for helpful discussions on the various dynamical processes that misalign disks; these conversations occurred at the Kavli Institute for Theoretical Physics (KITP) and were supported by grant no. NSF PHY-2309135.
This work was also supported by NASA through grant no. 23-XRP23 2-0148.
JPW thanks NASA and the NSF for support of basic research that actually does help make America great.
MB has received funding from the European Research Council (ERC) under the European Union’s Horizon 2020 research and innovation programme (PROTOPLANETS, grant agreement No. 101002188).
This paper is based on observations collected at the European Organisation for Astronomical Research in the Southern Hemisphere under ESO programme 1104.C-0415(D) (DESTINYS Large Programme).
GL acknowledges support from PRIN-MUR 20228JPA3A and from the European Union Next Generation EU, CUP:G53D23000870006.
This paper makes use of the following ALMA data: ADS/JAO.ALMA\#2018.1.00564.S. ALMA is a partnership of ESO (representing its member states), NSF (USA) and NINS (Japan), together with NRC (Canada), NSTC and ASIAA (Taiwan), and KASI (Republic of Korea), in cooperation with the Republic of Chile. The Joint ALMA Observatory is operated by ESO, AUI/NRAO and NAOJ. 
We thank the VLT and ALMA staff astronomers for making these observations possible.
\end{acknowledgments}

\begin{contribution}
JPW carried out the analysis and wrote the manuscript. MB initiated the project and provided feedback throughout. CG coordinated the SPHERE observations and reduced the data. GL provided theoretical input on the interpretation of the modeling results. MV helped with early stages of the project.
\end{contribution}

\facilities{VLT(SPHERE), ALMA}
\software{galario \citep{galario}, eddy \citep{eddy}, radmc3d \citep{radmc3d}, emcee \citep{emcee},
astropy \citep{2013A&A...558A..33A,2018AJ....156..123A,2022ApJ...935..167A},
matplotlib \citep{Hunter:2007}, scipy \citep{2020SciPy-NMeth}, scikit \citep{scikit-learn},
PyTorch \citep{PyTorch}.}

\appendix
\section{Model grid images}
We present a gallery of simulated SPHERE images from the model grid as in Figure~\ref{fig:representative_images} but where we change one parameter at a time (two for the scale height variation) while keeping the others fixed close to the interpolated best fit values in Table~\ref{tab:mcmc}.

\begin{figure*}
\begin{center}
\includegraphics[width=\textwidth]{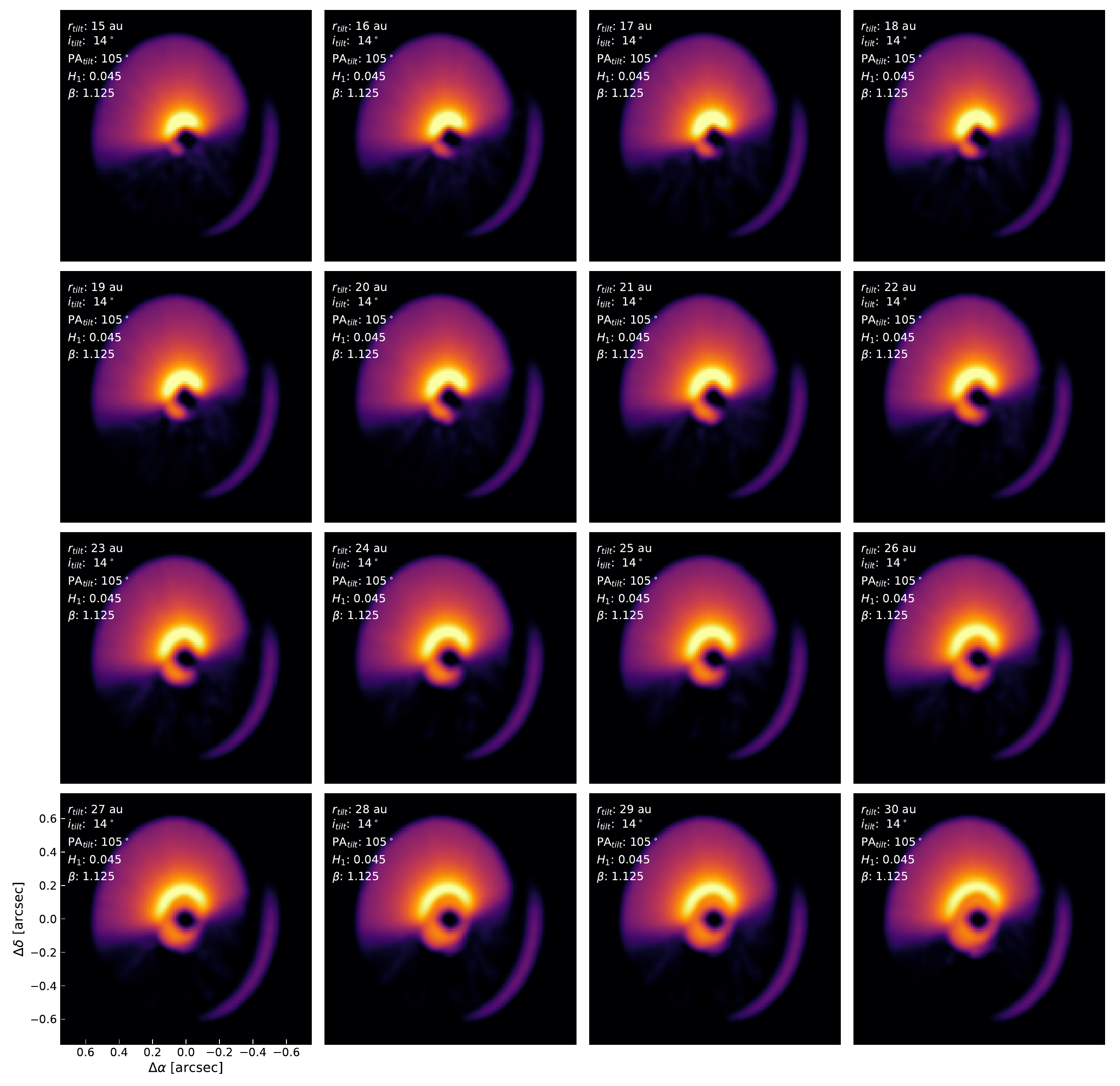}
\end{center}
\caption{Variation of inner disk radius, \Rtilt.
\label{fig:gridplot_radius}}
\end{figure*}

\begin{figure*}
\begin{center}
\includegraphics[width=0.7\textwidth]{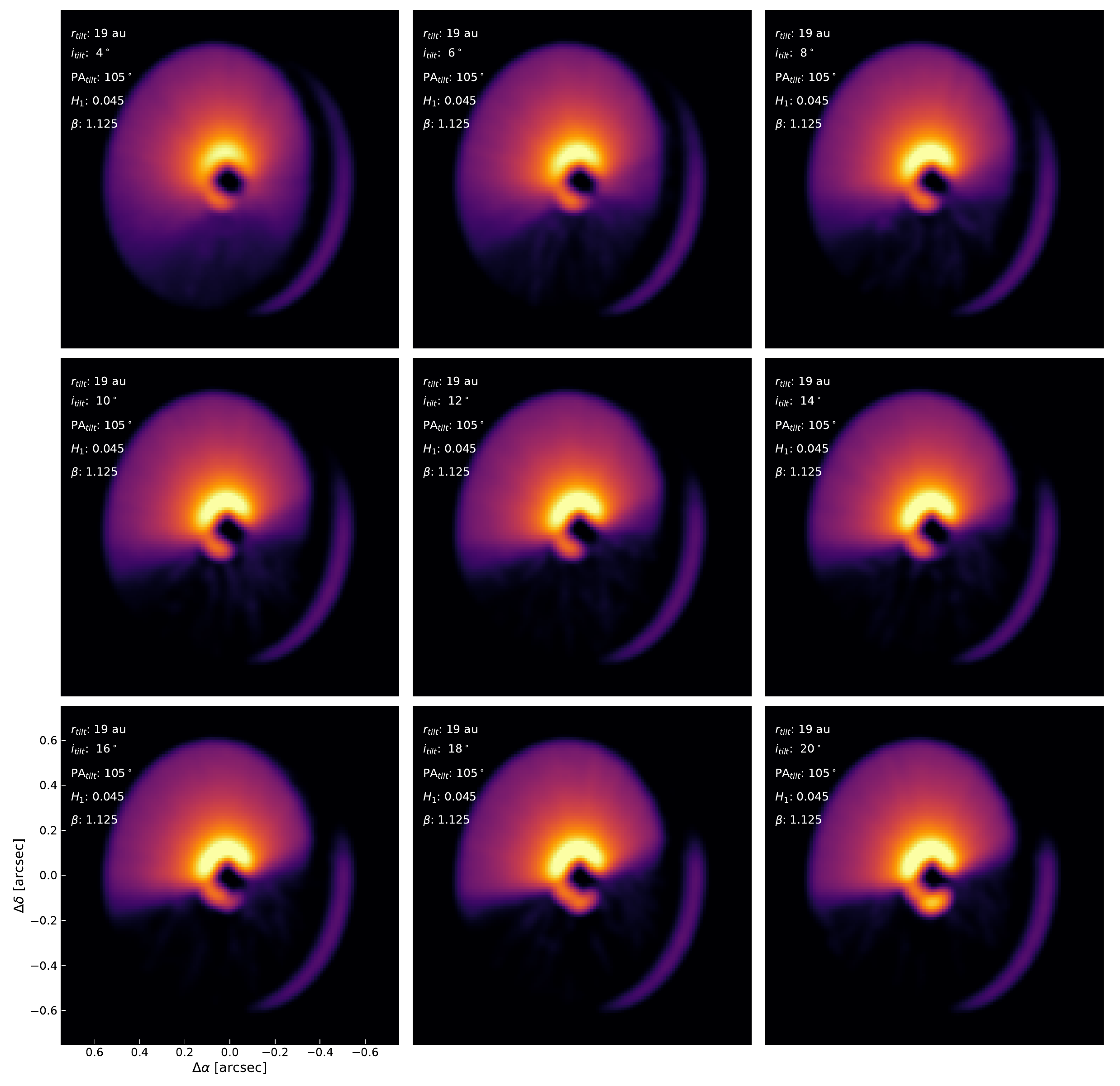}
\end{center}
\caption{Variation of inner disk inclination, \itilt.
\label{fig:gridplot_inclination}}
\end{figure*}

\begin{figure*}
\begin{center}
\includegraphics[width=\textwidth]{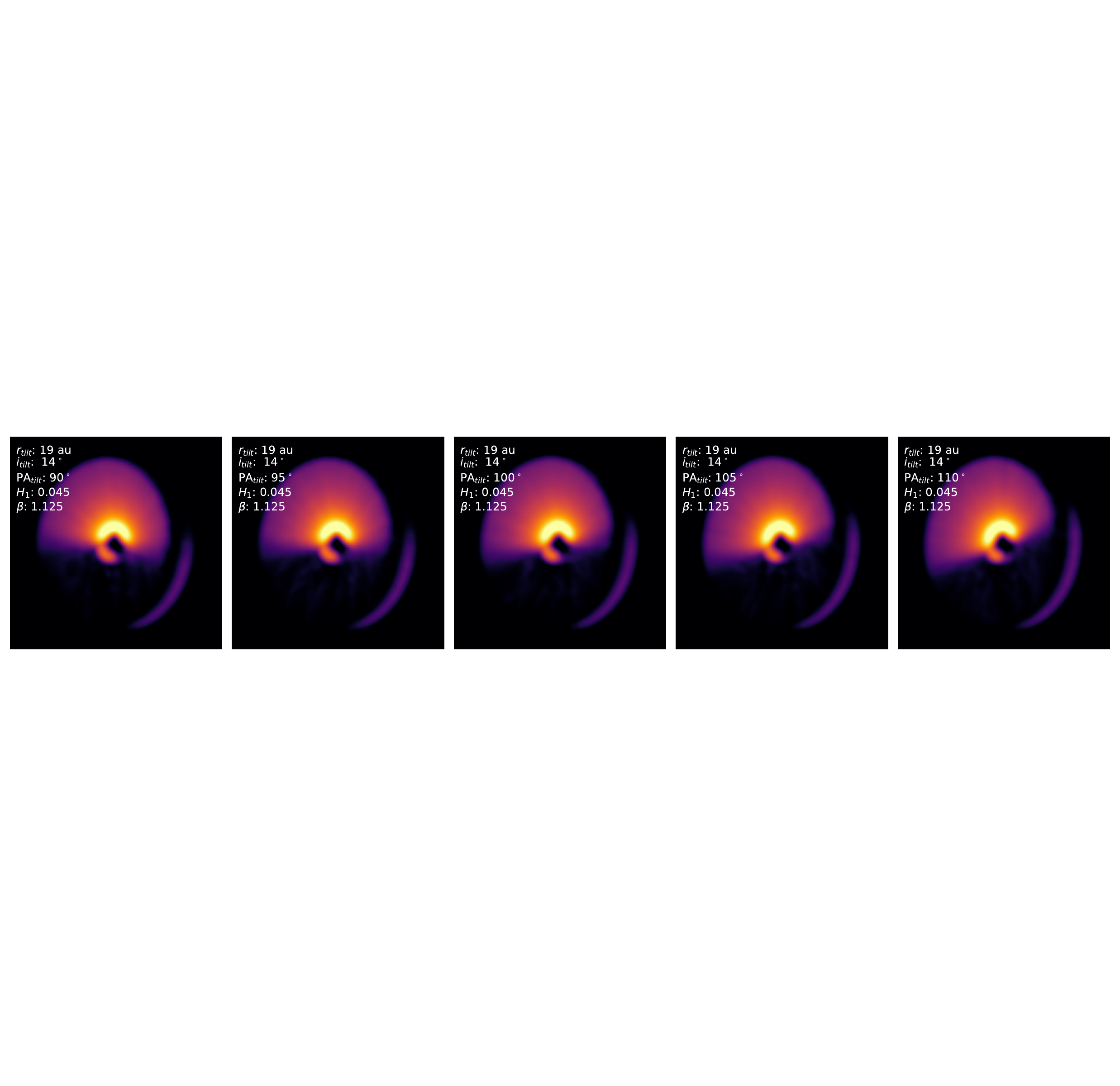}
\end{center}
\caption{Variation of inner disk position angle, \PAtilt.
\label{fig:gridplot_PA}}
\end{figure*}

\begin{figure*}
\begin{center}
\includegraphics[width=\textwidth]{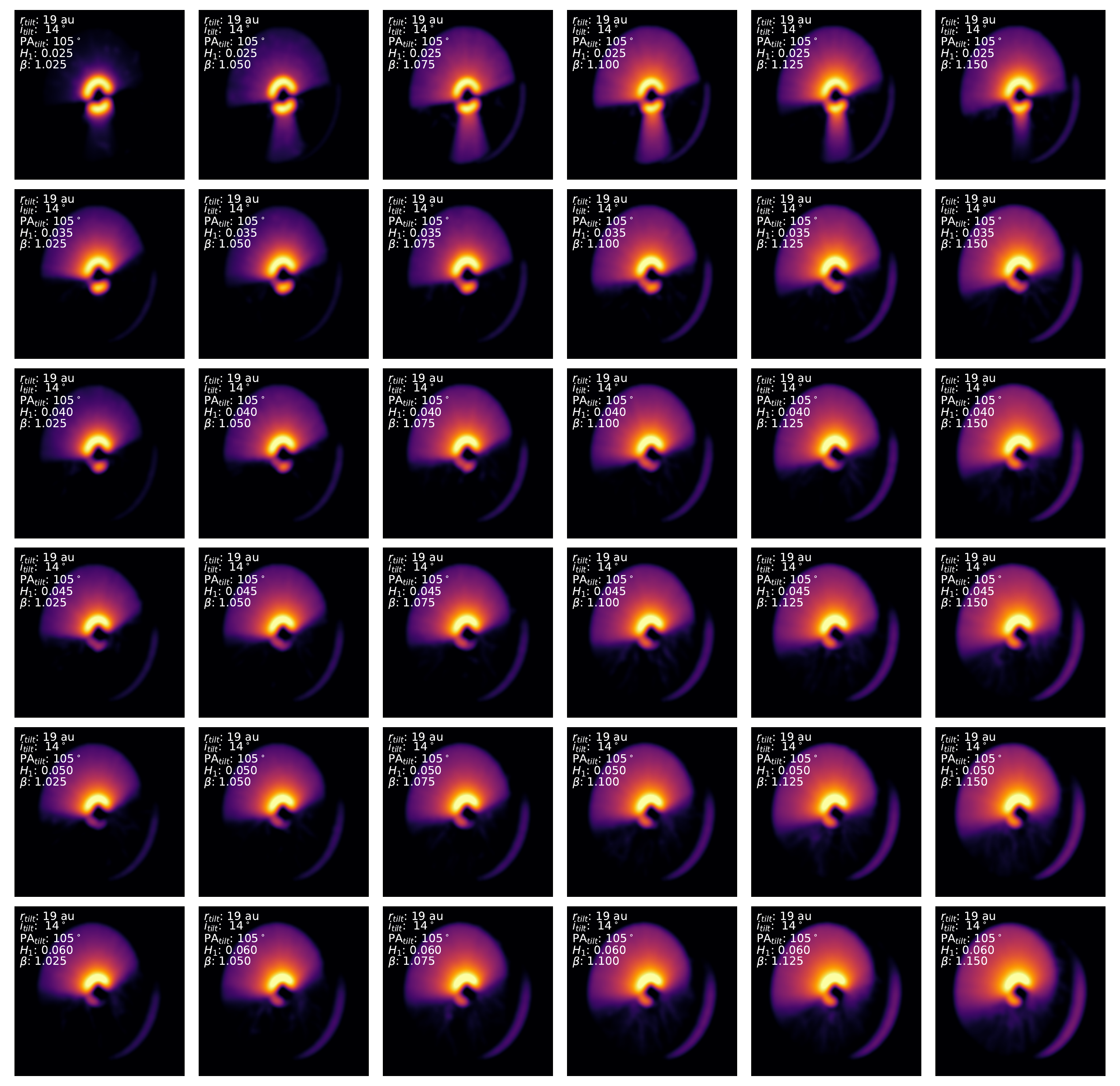}
\end{center}
\caption{Variation of scale height parameters, $H_1$ and $\beta$.
\label{fig:gridplot_flaring}}
\end{figure*}

\bibliography{refs, software, triangle}{}
\bibliographystyle{aasjournal}

\end{document}